# Learning Deep Models from Synthetic Data for Extracting Dolphin Whistle Contours

Pu Li[a], Xiaobai Liu[a]*, K. J. Palmer[a], Erica Fleishman[b], Douglas Gillespie[c]
Eva-Marie Nosal[d] , Yu Shiu[e], Holger Klinck[e], Danielle Cholewiak[f], Tyler Helble[g], and Marie A. Roch[a]*

[a] Dept. of Computer Science
San Diego State University
San Diego, CA, USA

[b] Dept. of Fish, Wildlife
and Conservation Biology
Colorado State University
Fort Collins, CO, USA

[c] Sea Mammal
Research Unit
Scottish Oceans Institute
University of St Andrews
St Andrews, Fife, Scotland

[d] Dept. of Ocean and
Resources Engineering
University of Hawaiʻi at
Mānoa, Honolulu, HI, USA

[e] Center for
Conservation Bioacoustics,
Cornell Lab of Ornithology,
Cornell University
Ithaca, NY, USA

[f] Northeast Fisheries
Science Center, National
Marine Fisheries Service,
National Oceanic and
Atmospheric Administration
Woods Hole, MA, USA

[g] US Navy
Naval Information Warfare
Center Pacific
San Diego, CA, USA

* corresponding authors:
xiaobai.liu/marie.roch
@sdsu.edu

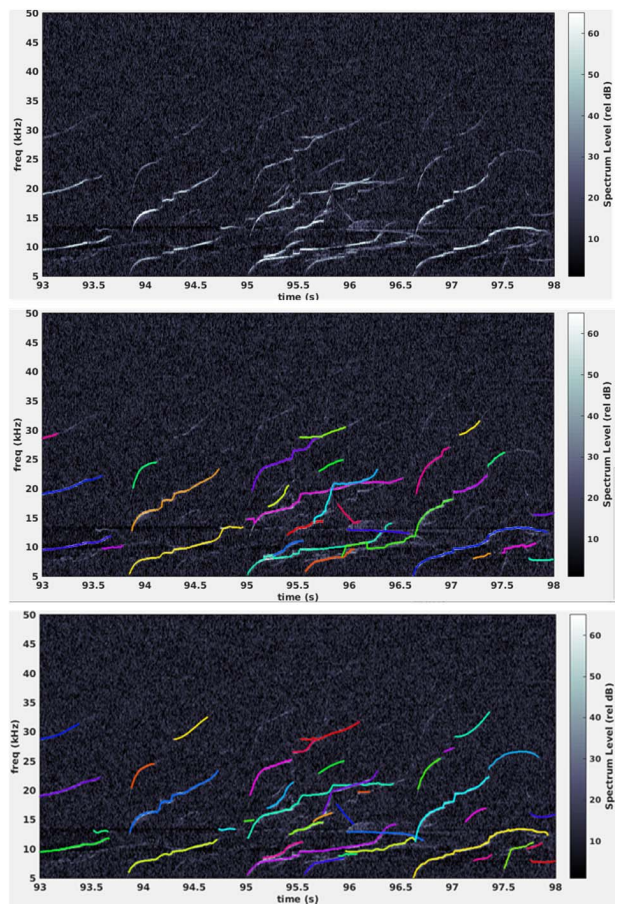

Fig. 1 Time-frequency spectrogram (upper panel) with whistles produced by common dolphins (*Delphinus* spp.). Randomly colored annotations of whistle contours produced by our system (middle) and a skilled analyst (lower).

*Abstract*—We present a learning-based method for extracting whistles of toothed whales (Odontoceti) in hydrophone recordings. Our method represents audio signals as time-frequency spectrograms and decomposes each spectrogram into a set of time-frequency patches. A deep neural network learns archetypical patterns (e.g., crossings, frequency modulated sweeps) from the spectrogram patches and predicts time-frequency peaks that are associated with whistles. We also developed a comprehensive method to synthesize training samples from background environments and train the network with minimal human annotation effort. We applied the proposed learn-from-synthesis method to a subset of the public Detection, Classification, Localization, and Density Estimation (DCLDE) 2011 workshop data to extract whistle confidence maps, which we then processed with an existing contour extractor to produce whistle annotations. The F1-score of our best synthesis method was 0.158 greater than our baseline whistle extraction algorithm (~25% improvement) when applied to common dolphin (*Delphinus* spp.) and bottlenose dolphin (*Tursiops truncatus*) whistles.

*Keywords—Whistle contour extraction, deep neural network, data synthesis, acoustic, odontocetes*

## I. Introduction

Many animals produce calls that may contain information on species' identity (e.g., [1]), movement (e.g., [2]), individual identity (e.g., [3]), behavior (e.g., [4]), or provide clues into density and abundance (e.g., [5]). Passive acoustic recorders collect audio data that can be analyzed to yield such information. The cost of data collection has fallen due to innovation driven by demand for consumer electronics, but processing the increasing volume of data is an ongoing challenge. Therefore, there is a need for tools and methods to extract detailed animal acoustic signals from continuous acoustic recordings.

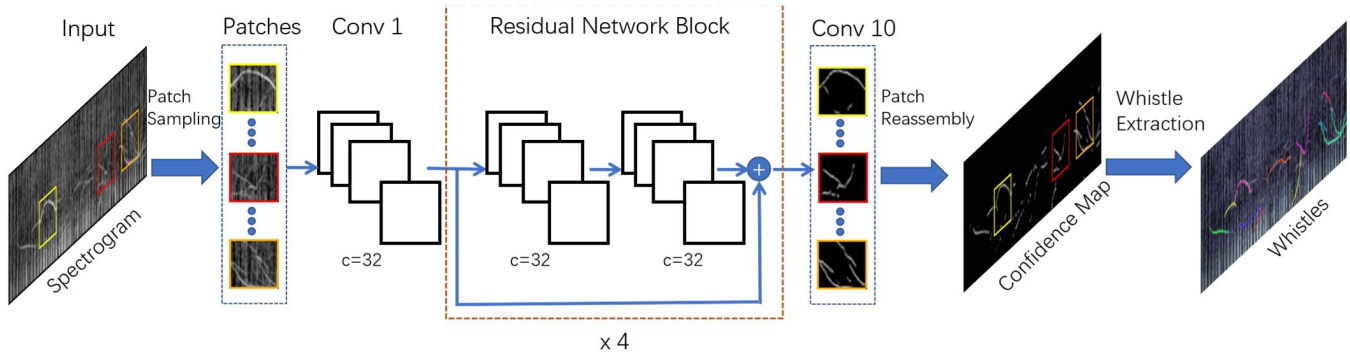

Fig. 2  Flowchart of our method. Boxes represent an incomplete set of the 100 ms by 6.25 kHz patches that tile the spectrogram (not to scale). Patches are passed to a 10-layer network with four residual blocks resulting in confidence maps that predict whistle energy are produced. The maps are reassembled and can be used with search algorithms to extract whistles.

Our objective was to develop automated methods to extract whistles from an environment with many confounding sound sources. To illustrate, we applied our method to an input spectrogram (Fig. 1 upper panel) to produce whistle contours (middle panel). Contours represent the time-frequency traces of odontocete whistles. Ecological, behavioral, and communication questions, such as individual identity and population density, can be answered with acoustic recordings of animals. In many cases, doing so relies on detailed information about individual animal call attributes including duration, frequency range, and frequency modulation. Therefore, in many contexts, it is insufficient to simply know that animals are present and calling.

Contour detection can be challenging due to spatial and temporal differences in ocean soundscapes, including varying sea state; precipitation; animal behavior; whistles that overlap in time and frequency; non-target biotic sounds, such as dolphin echolocation clicks or the impulsive signals of snapping shrimp; and abiotic sounds, such as those made by vessels.

In section II, we review existing methods for extracting whistle contours and related problems. Among these methods, learning-based approaches can model the statistical properties of whistle contours directly and often perform better than other methods. However, it is expensive and time-consuming to manually annotate whistle contours in time-frequency spectrograms, and doing so requires expertise in bioacoustics. Moreover, training a modern machine-learning model, e.g., a deep neural network, often requires many more annotated samples than classical models. Therefore, we investigated learning-based methods that can leverage small numbers of human annotations, or no annotations at all, to predict whistles.

We aimed to generate a pointwise, two-dimensional confidence map to represent whistle contours in time-frequency spectrograms of varying durations. Each element of the map represents the likelihood that a corresponding time-frequency point in a spectrogram belongs to a whistle contour. Similar to edges or boundaries in natural images, short segments of whistle contours often have primitive (simple) shapes, such as a crossing between two whistles or a frequency-modulated sweep. For example, in Fig. 1 middle panel, the light blue and purple whistles overlap between 95.5 and 96 s.

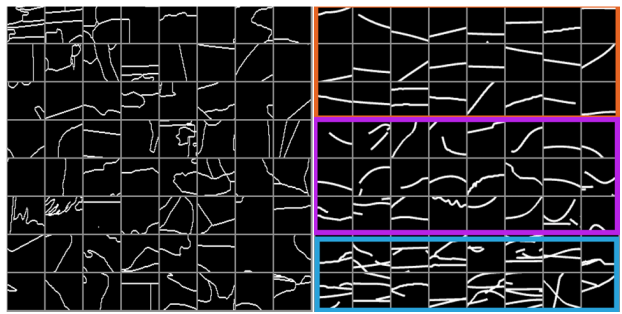

Fig. 3  Sample of shapes from which the implicit dictionary was derived. Left: shapes in natural images, extracted from the boundary maps in the Berkeley Segmentation Dataset 500. Right: shapes derived from whistle annotations in time-frequency spectrograms; linear, curved, and crossed whistles are highlighted by orange, magenta, and blue boxes, respectively.

Classical computer vision methods, e.g., primal sketch [6] or textons [7], employ a set of primitive shapes to parse natural images or videos. Rather than directly learning a dictionary of such primitive shapes, we employed a deep convolutional network to implicitly learn the whistle information from small patches of natural or synthetic spectrograms and a label set that indicated where whistles occurred within each patch. During testing, we partitioned the spectrogram into small patches and assembled the confidence map predictions into a larger map, which we then processed with an existing, graph-based whistle extractor to measure the performance of our system and provide an equivalent comparison with other methods (Fig. 2) [8].

The performance of the proposed method largely depends on the quality and quantity of patch-shape pairs used to train the deep network. We found in our experiments (section IV) that the performance of our method decreased substantially as the number of training samples decreased. To overcome this, we developed a comprehensive learning-from-synthesis approach to train the network from synthetic data. We sampled small time-frequency patches from spectrograms of ocean sound without whistles. Into these, we blended primitive, whistle-like shapes, permitting the learning of whistle characteristics in diverse sound environments. We examined multiple methods that collect primitive shapes from different sources, including whistle annotations in spectrograms and edge annotations in natural images. We also used the edge detections of classical computer vision methods (e.g., Canny's edge detector [9]) to synthesize data, which allowed us to extract whistles without using any human annotations. We used these automatically generated spectrogram patches and their shapes to train the deep network. Empirical studies of these data-generation methods suggested that our method is capable of achieving high accuracy while substantially reducing the amount of human effort required.

## II. RELATION TO PREVIOUS WORK

### A. Whistle Contour Extraction

Previous researchers used different methods to extract delphinid whistle contours. Trajectory-search methods seek peaks in the spectral energy of short consecutive segments and stitch neighboring peaks together on the basis of a trajectory estimate ([1, 8, 10]). Other work has examined local context, performing ridge regressions [11] or building on ridge regression maps by using energy minimization algorithms to find contours followed by a final classification to reduce excessive numbers of false positives [12].

Probabilistic frameworks are an alternative approach to extracting whistle contours. Examples of this approach include hypothesis tests of spectrogram region distributions [13], Bayesian inference [14], Kalman filters [15], particle filters [8, 16], and multi-target tracking [17].

Neural networks have been used to extract tonal information in human speech and music tasks [18, 19], and usually perform better than probabilistic methods when the number of annotations is sufficient. The same methods may be applicable to problems in marine acoustics. However, the latter application has not been tested, in part because there are orders of magnitude fewer annotated data for marine mammal tonal calls than human speech and music. These techniques are promising and have the potential to significantly improve the efficacy of passive acoustic monitoring and animal conservation efforts.

### B. Deep Models

Convolutional neural networks are effective for pixel-wise classification in computer vision methods, such as image contour detection [20-23] and semantic segmentation [24]. These neural networks implicitly leverage the context around each pixel to predict labels. Our aim is analogous: to predict the location of whistle time-frequency nodes on the basis of the local intensity and shape of the whistle contours in spectrograms.

### C. Dictionary Learning

Dictionary learning [25] is a strategy for discovering a set of vectors, or atoms, that can be used to reconstruct elements of a data set. The well-known $k$-means algorithm [26] is a simple example of such learning, with the $k$ mean vectors serving as the atoms of the dictionary. Dictionary learning has been applied to tasks including image super-resolution [27], face recognition [28], image denoising [29], and visual tracking of objects [30]. Traditional methods usually build a dictionary of bases (atoms) and then use the dictionary to reconstruct the input image. The dictionary often is explicitly specified through hand-crafting or discriminative training. In this work, we built an overcomplete dictionary of primitive whistle shapes to represent time-frequency spectrograms. During testing, we sampled multiple small-size time-frequency patches from the input spectrogram and employed a deep neural network to retrieve the primitive whistle shape of each patch.

### D. Learning from Synthesis

The power of deep convolutional neural networks to detect and classify objects is limited when there are insufficient quantities of annotated data. Data synthesis, or the use of algorithms to develop artificial training examples, can help to fill this gap. Data synthesis differs from data augmentation, which is a transformation of an existing instance. Data synthesis has increased performance in the fields of text localization [31] and instance detection [32]. In this work, we examined multiple approaches for generating time-frequency whistle patches to augment or entirely replace training data. We also studied a network learning algorithm that allowed us to adaptively synthesize the most important samples over training steps.

## III. APPROACH

Our method for extracting whistle contours in time-frequency spectrograms has three major steps: (i) sample patches from the input spectrogram, (ii) feed each patch into a neural network to obtain its corresponding confidence map, (iii) aggregate patch confidence maps to obtain an overall contour confidence map the same size as the original input spectrogram (Fig. 2). We used graph-search [8] to extract discrete whistle contours from the aggregated confidence map. The network learned a representation of the contour patches. Effective training of such a network usually requires hundreds of thousands of annotated examples. The quality of annotations largely relies on expert knowledge of the sounds that the target taxonomic group produces, and obtaining annotated data is time and labor intensive. Limiting or eliminating the need for humans to annotate spectrograms while maintaining performance levels would be a major advance in learning-based whistle detection.

### A. Data and Signal Processing

We used a subset of the annotated delphinid data from the Detection, Classification, Localization, and Density Estimation (DCLDE) 2011 workshop [33]. We standardized the recordings (192 kHz sample rate, 16 or 24 bit quantization) to 16 bit quantization. We restricted our analysis to two taxa: common dolphins (*Delphius capensis* and *D. delphinus*) and bottlenose dolphins (*Tursiops truncatus*). Common dolphins were the more challenging taxon due to their tendency to aggregate in large numbers with many simultaneous whistles, resulting in numerous whistle crossings. We selected 20 audio sequences that had time-frequency contour annotations created by analysts with the aid of an interactive toolkit [8]. We used 14 of these recordings for training and the remaining ones for test. This resulted in a total of 7,161 whistle contours in the training sequences and 911 whistle contours in test sequences.

We created log-magnitude spectrograms from discrete Fourier transforms of 8 ms Hamming windowed frames (125 Hz resolution) advanced every 2 ms. We restricted the dynamic range of the resulting spectrogram to a floor of 0 and a ceiling of 6 (an intensity range of 0 to 120 dB rel., uncalibrated and unadjusted for frequency bin width) on the basis of empirical results. We normalized spectrogram values to the interval [0, 1]. We then frequency-limited these data between 5 and 50 kHz (361 frequency bins), the range of most delphinid whistles and their harmonics.

We partitioned spectrograms into 100 ms by 6.25 kHz (50 x 50 time-frequency bins) patches. In the training data, patches were formed by tiling the spectrogram into patch-sized regions that overlapped by 50%. Patches were split into whistle-positive and whistle-negative groups. There were 74,112 positive patches and we randomly selected an equal number of negative patches. We used these data, or random samples thereof, to train the deep contour model.

We also examined whether the model could learn from alternative data sources (III.C). To do so, we examined three different sources: frequency contour information provided by the DCLDE 2011 annotation data (no spectrograms), a set of analyst-annotated image edge annotations from the Berkeley Segmentation Dataset (BSDS 500) [34], and a set of edges generated automatically from the BSDS 500 images by a Canny edge detector [35]. Regardless of the source, we injected the annotations into whistle-absent spectrogram patches.

We used a similar processing chain to process data for testing, with additional steps as outlined below.

*B. Deep Representation of Whistle Contours*

In the experiment without data synthesis, we used analyst ground-truthed whistle (WGT) annotations of real data to determine the model's capacity to learn to represent small sections of whistles when training data are relatively abundant. Training data consisted of the 148,224 positive and negative spectrogram patches and corresponding 50 x 50 whistle confidence labels, which we set to 1 in each time-frequency node that contained a whistle and to zero otherwise. We trained a fully convolutional deep network [22] with stochastic gradient descent from these data, the architecture of which is detailed in section III.E.

When using the trained network to extract whistles, we partitioned the spectrograms of the test dataset into non-overlapping patches. The network produced confidence maps that predicted the location of whistle energy in the input spectrogram patch. We reassembled these confidence maps into a spectrogram-like structure that we passed to a peak processing algorithm [8] which produced a set of hypothesized whistles. The choice of post-processing algorithm was one of convenience, and we expect that other peak-based assembly algorithms will yield similar improvements when peaks are selected in a more reliable manner.

Similar deep models have been used to detect edges or boundaries in images, and were much more accurate than traditional edge detection methods when applied to standard benchmarks (e.g., BSDS 500 [34]). The learning in such deep models, however, requires thousands of human-annotated image-contour pairs.

By restricting our analysis to small spectrogram patches and eliminating large segments of the spectrogram in which whistles were absent, we substantially reduced class imbalance. The convolutional filters were no larger than 10 ms by 375 Hz (see III.E for details), and the network had a receptive field of 46 ms by 2.875 kHz. Whistle fragments in the resulting receptive field had relatively simple shapes. Longer whistle fragments would have had a more complex distribution that is likely to present greater challenges to model in a larger receptive field. Small patches and receptive fields allowed us to exploit multiple methods for generating a sufficient volume of synthetic training data to learn to predict confidence maps on the basis of little to no whistle data.

*C. Synthesizing Training Data*

We developed methods to generate synthetic training data to minimize human effort and increase the volume of training data. We injected primitive shapes (Fig. 3) into 100 ms by 6.25 kHz spectrogram patches of whistle-absent recordings, which provided realistic examples of ocean ambient sounds (Fig. 4). In these experiments, we selected these examples from our whistle-negative patches, but one could use this method to synthesize data for a novel recording environment.

We explored two methods for data generation. First, we used contours collected from the computer vision domain to create synthetic data. These were either analyst-annotated boundaries of images in the BSDS 500 data [34] or edge segments that we detected automatically with Ding et. al.'s modifications to the Canny edge detector [35]. These contours contain many shapes similar to very short segments of delphinid whistles, although scenes containing anthropogenic landmarks (e.g., buildings) can have edges with sharp corners or vertical excursions that are not typically present in delphinid whistles. Second, we used samples of the human-analyst annotated whistle contours in the DCLDE 2011 data. We created synthetic data, injecting these accurately captured whistle shapes into spectrograms of ocean sound data.

We created binary label masks ($Y$) for each synthetic example (1 indicated whistle presence) and convolved them with a random Gaussian filter ($G$) to blur each binary mask. We aligned the blurred binary contour mask with a whistle-absent spectrogram patch to generate whistle-present training samples. We obtained the synthesized sample $X'$ via

$$X' = X + \alpha(Y * G), \qquad (1)$$

where X is a background spectrogram patch, α is a random intensity parameter drawn from a uniform distribution over [0.03, 0.23], and $*$ is the convolution operation. This results in signal strengths varying over 24 dB. The signal-to-noise ratio (SNR) of these signals was dependent on the noise levels in the spectrogram patches into which they were blended and the randomized signal intensity, and these values were representative of typical SNR values in real data. Our sample synthesis method reduced or, in some cases, eliminated the need for humans to annotate whistle contours (Fig. 4).

*D. Simultaneous Learning and Sample Synthesis*

Traditional learning-from-synthesis methods usually generate many samples to ensure effective learning. However, it is unclear how to determine the minimum number of synthesized samples for a particular application. Moreover, this strategy assumes that all examples are equally important and generates a homogeneous set of synthesis samples. Providing equal weight to examples that are easy and difficult to learn is inefficient, causing the network to expend more time considering examples that it already predicts well. Biasing the synthesis process to produce samples in proportion to their difficulty should increase the effectiveness of the learning process.

We developed an integrated approach that adaptively generates samples while training the network. This process focuses on those spectrograms or contour segments that were not well modeled by the network. We based our two-stage learning algorithm on the standard mini-batch gradient descent method. In the first stage, we restricted training contours and sound patches to a small subset (6.25%) of the available data, from which we generated synthetic whistle-positive examples. We also included an equal number of randomly selected negative examples. We used these positive and negative samples to train an initial network. As we describe in the experiment section (IV), use of this small amount of data with standard training techniques (no synthesis) yielded confidence maps with low recall (Fig. 5).

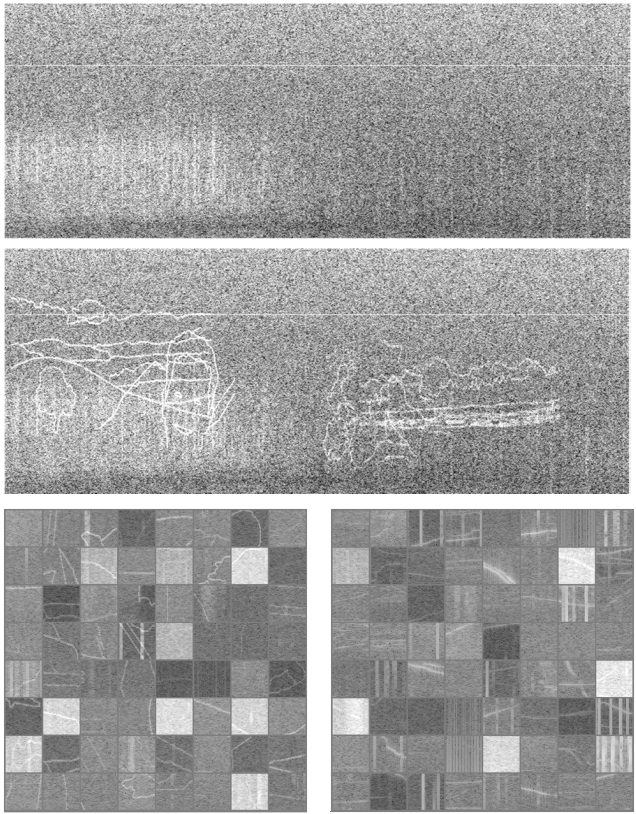

Fig. 4 Example of data synthesis. Top: background spectrogram. Middle: spectrogram overlaid with natural image edges. Bottom: synthetic patches from BSDS 500 labels (left) and DCLDE 2011 whistle contours (right).

The second stage was an iterative process with multiple training epochs. At the beginning of each epoch, we used the network weights from the previously completed epoch to produce confidence maps of the validation data. Validation data consisted of positive and negative spectrograms containing the non-synthesized data that were available to the algorithm (the 6.25% subset of the annotations). We computed the recall rate of the confidence maps, which allowed us to identify which positive examples were more difficult for the current network to classify. We assessed recall for a sample $z$ as

$$R(z) = \frac{\sum\sum(I_{\geq 0.5}(C_z) \cdot Y) + \epsilon}{\sum\sum(Y) + \epsilon}, \qquad (2)$$

where $C_z$ is the confidence map predicted from $z$, $I$ is an indicator function thresholded at 0.5 that binarizes $C_z$, $Y$ is the label mask, · is the element-wise (Hadamard) multiplication operator, and $\epsilon$ is a small value to prevent division by zero. The double summation is over the rows and columns of the confidence map. We chose to use recall rates instead of precision rates to retain as many contour candidates as possible during the early stage of the learning process.

We next used probabilistic selection to generate a new set of training data with roughly equal numbers of positive and negative examples. Let $p(z)$ be the probability that training sample $z$ is selected. Let $R(z, t)$ be the recall for sample $z$ in the $t^{th}$ epoch. Then the probability of selecting a negative ($N$) or positive ($P$) sample $z$ at epoch $t$ is

$$p(z) = \begin{cases} 0.5 \frac{1}{|N|}, & z \in N \\ 0.5 \frac{1 - R(z, t-1)}{\sum_{\zeta \in P}(1 - R(\zeta, t-1))}, & z \in P \end{cases} \qquad (3)$$

which biases selection of positive examples towards those with lower recall. We used the whistle annotations associated with selected positive examples to synthesize new examples. This enabled the network to observe the difficult pattern in varied contexts. This adaptive learning process can synthesize difficult samples over training epochs and improves performance.

In summary, the key components of our learning algorithm include sampling difficult examples of primitive whistle shapes from $p(z)$, synthesizing new samples for the selected shapes, and updating the network parameters on the basis of the new synthesized samples. In this way, network learning and sample synthesis are alternated until convergence.

*E. Implementation*

**Network Design.** Our network has 10 convolutional layers and each hidden layer has 32 channels. We added batch normalization layers behind all convolutional layers on residual branches. We used parametric rectified linear unit (Prelu) layers [36] as nonlinear layers behind the first batch normalization layers on the residual branches. The first and last convolutional layers had a filter size of 5 and zero padding size of 2; the remaining convolutional layers had a filter size of 3 and zero padding size of 1. The eight layers in between formed four residual blocks [37]. All layers had a stride size of 1. This simple architecture had a first-layer convolutional filter of size 50 ms by 3.125 kHz region in the time-frequency domain. Subsequent layers had smaller 25 ms by 0.625 Hz filters. We did not use pooling in this architecture, which makes predictions on individual time-frequency nodes.

**Extracting Whistles From Confidence Maps.** In peak processing whistle extraction methods, extraneous sound sources in the spectrograms frequently produce false positive peaks. When there is enough structure in the extraneous above-ambient peaks or the distance between peaks is short, the extraneous peaks can lead to false-positive whistle contours. Similarly, a high number of missed peaks can lead to false negatives. A more-reliable method of identifying peaks on the basis of contextual cues would dramatically improve the robustness of such algorithms. We replaced the spectrogram input with a confidence map of the probability that each time-frequency node is part of a whistle contour.

We employed an existing graph search algorithm [8] that organizes time-frequency peaks into a graph to test the reliability of our system's peak estimation and its capacity to improve peak tracking algorithms by exploiting contextual cues in the deep network. We applied this post-processing step to the test data after the confidence map patches were reassembled into a spectrogram-like structure and whistle energy was identified with a threshold $\tau$ of 0.5.

At each time step, whistle energy in the confidence map either initiates a new graph or extends terminating nodes of existing graphs. Extension is chosen when the new peak is along a reasonable trajectory predicted by a low-order polynomial fit of the graph path near a terminating node. Graphs can bridge gaps in detected peaks up to a user-established time interval, reducing the impact of missed peak detections. When a peak is added to a pair of graphs, the graphs are merged. Graphs that have not been extended within a specified time are considered closed. Internal nodes of closed graphs with multiple inputs and outputs represent potential whistle crossing points. We used an analysis of forward- and backward-predicting polynomial fits to determine which branches of the graph should be placed together in the hypothesized whistle contours.

IV. EXPERIMENTS

*A. Evaluation Protocol*

**Metrics.** To assess the quality of the confidence maps before whistle contour extraction, we employed the BSDS 500 benchmark tools and protocol [34] to calculate precision and recall. We thinned our ground-truthed confidence maps to 1-pixel wide and compared them with a predicted confidence map binarized by 30 thresholds between (0, 1). We applied all default parameters in the evaluation tools.

We used three metrics to evaluate and compare the quality of discrete whistle contours predicted by the complete contour extraction system: (i) recall, the percentage of validated whistle contours that were detected, (ii) precision, the percentage of detections that were correct, and (iii) F1-score, or the harmonic average of precision and recall. We determined success or failure of whistle extraction by examining the set of expected analyst detections as described in [8]. Briefly, for each analyst-annotated whistle contour, we checked whether any of the detections overlapped in time. If so, we examined whether each overlapping detection matched the analyst annotation. We considered that a match occurred if the average deviation in frequency between the two contours was < 350 Hz and the analyst detections were ≥ 150 ms in duration, with a signal to noise ratio ≥ 10 dB over at least a third of the whistle. When detection of a whistle was not expected (too short or low intensity), we discarded the matched detections, and they did not contribute to the metrics. We counted unmatched detections as false positives. Due to the multiple decision criteria, we evaluated the algorithm at the single operating point used in [8] and DCLDE 2011.

**Experiments.** We implemented five variants of our whistle contour extraction method to quantify the contributions of the deep contour model, data synthesis, and recall-guided learning. All but one of these variants used human or machine-detected annotations to synthesize training samples. We compared the variants to the baseline method [8]. We used a fully convolutional network, trained on the complete set of ground-truthed whistle data (WGT), to assess the learning capacity of the deep contour model. The remaining variants trained the network exclusively with synthesized data. EdgeGT used all analyst-annotated edges from the BSDS 500 data [34]. The training data in EdgeCanny were not analyst annotations, but automatically generated edges from the BSDS 500 images. The remaining variants were trained with data synthesized from delphinid whistle annotations. μWGT employed a small amount of whistle contour time-frequency information from the 2011 DCLDE data (we did not use audio data containing real whistles) to synthesize spectrogram-contour samples. For μWGT, we simulated limited annotated data by randomly selecting 6.25% (4,632) of the patches in WGT. The analyst contours associated with the selected patches were used to synthesize samples. μWGT-RG extended μWGT with the our recall-guided learning-with-synthesis algorithm. We used the confidence maps generated by the above networks as inputs to call the graph-search method [8] to generate whistle contours (Fig. 2).

We used Pytorch [38] for all our experiments. We trained all network variants with the Adam optimizer [39] with 600 000 iterations. We used Charbonnier loss [40] for network gradient calculation:

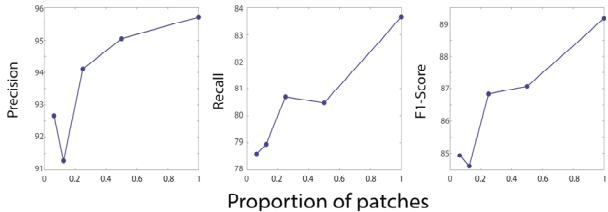

Fig. 5   Confidence map metrics that were based on 6.25, 12.50, 25.0, 50.00, and 100% of the 148,224 training patches used in WGT.

$$\text{Loss}(\hat{y}, y) = \sqrt{||\hat{y} - y||_2^2 + \varepsilon} \quad (4)$$

where $||\cdot||_2^2$ is the squared L2 norm, $\hat{y}$ is the network output, y is the ground-truthed data, and $\varepsilon$ is $1\times10^{-3}$. We used an initial learning rate of 0.001, and decayed the learning rate by a factor of 0.1 every 250k iterations and initialized the networks with the Kaiming Normalization [41]. The μWGT-RG network was initialized to the final μWGT network.

*B. Analyses and Results*

**Number of Training Samples.** We first evaluated how the quantity of annotated whistle contours affected the prediction of confidence maps. System performance increased as the volume of training data increased (Fig. 5). Training an effective deep contour model required a large number of annotated samples. The largest performance increase occurred as the selected proportion of training data approached 25%, suggesting that the data have some redundant patterns. Our evaluation suggested that it is valuable to synthesize many training samples.

**Prediction Execution Time.** We tested and ran our algorithms on a workstation with an NVIDIA GTX 1080Ti GPU. Training required about 8 hours, with some variation among model variants. Testing 1 s of data required about 10 ms for network prediction plus 300 ms for the graph search. Our method can extract whistle contours approximately 3 times faster than real-time.

**Quantitative Results.** We evaluated our system against portions of the DCLDE 2011 data [33] that contained 911 test whistles meeting the SNR and duration selection criteria described above. There were 354 bottlenose dolphin and 557 common dolphin whistles.

For confidence map evaluation (Fig. 6), WGT, the method designed to test model capacity with larger amounts of training data, dominated the precision-recall curves of the data synthesis methods. The synthesis methods based on delphinid whistle contours, μWGT and μWGT-GT, generally performed better than EdgeGT and EdgeCanny, which relied on image edges. EdgeGT and EdgeCanny, however, had higher precision than the other synthesis methods at lower recall.

For the two-stage whistle extraction system, we compared our previous work [8] to systems that used the same time-frequency peak processing rules, but relied on peak predictions from our confidence maps (threshold 0.5, Table I). The scores of the graph search algorithm were lower than those reported in [8] because we focused on a difficult subset of the data. The average F1-score of the fully trained model, WGT, was 0.250 greater than that of the graph search method without the use of confidence maps.

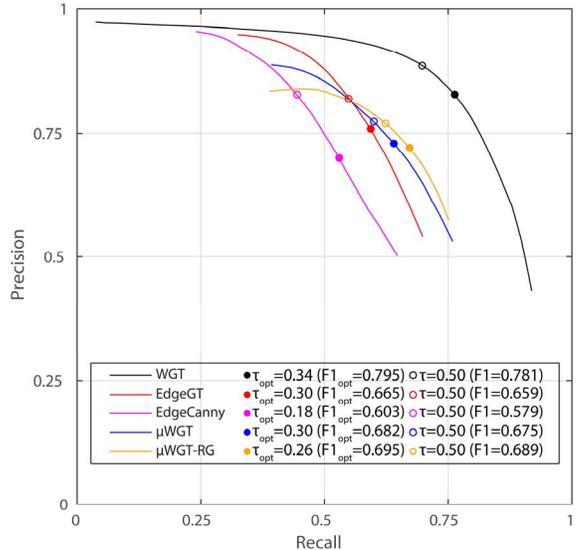

Fig. 6   Performance (precision and recall curve) of the five deep models predicting time-frequency nodes containing whistle energy. Circles indicate operating points (at threshold τ) along the curve. Filled circles: optimal F1 performance. Open circles: operating point used when extracting whistles with the post-processing peak assembly algorithm.

All variants of the models that used synthesized data outperformed the baseline version of the graph search method. The F1-score results of EdgeCanny suggested that networks can outperform existing techniques by a considerable margin of 0.088 (~14% improvement) without the use of analyst annotations. The use of image edges is particularly useful when no whistle contours are available. The results of μWGT and μWGT-RG demonstrated that synthetic whistles based on real whistles improved the F1 scores, and that recall-guided

TABLE I.   CONTOUR EXTRACTION PERFORMANCE

| 911 ground truth whistles | | | |
|---|---|---|---|
| **Contour extractor** | *Precision* | *Recall* | *F1-score* |
| Graph-Search | 0.634 | 0.633 | 0.634 |
| WGT | 0.956 | 0.822 | 0.884 |
| Synthesis-based experiments | | | |
| EdgeGT | 0.913 | 0.638 | 0.751 |
| EdgeCanny | 0.900 | 0.603 | 0.722 |
| μWGT | 0.900 | 0.673 | 0.770 |
| μWGT-RG | 0.924 | 0.693 | 0.792 |

learning could produce further improvements.

**Qualitative Results.** Fig. 7 represents a particularly challenging acoustic scene for whistle extraction systems. These data contained delphinid whistles and signals from a shipboard echosounder. Analyst annotations (top row) are compared with the baseline graph search (middle row) and the same graph search with peaks identified by WGT generated confidence maps (bottom row). This sequence, like many others in the DCLDE 2011 data [33], included changes in sound regime, anthropogenic signals, and other types of segments and reconstructed confidence maps from them. Because there was no attempt to minimize the number of implicit representations, we expect that an implicit representation of the dictionary will be overcomplete. The success of such a dictionary-based method depends on the quality of the retrieval and reconstruction procedure. Alternatively, the method simply may be learning to leverage the local context to improve the prediction of when time-frequency nodes are part of a larger whistle structure. Regardless, it is clear that the network learns to find whistle elements while ignoring distractors such as echolocation clicks (Fig. 2; clicks appear as vertical lines in the input spectrogram and are absent in the confidence map).

## V. CONCLUSIONS

We introduced a learning-based method for extracting whistle contours in time-frequency spectrograms. Our method learned to recognize local shape in whistle contours and reconstruct confidence maps in a manner analogous to dictionary learning. Our method used a small number of human or machine-generated annotations (whistle contours or edges in images) to learn to predict whistle contours, and outperformed a representative peak-tracking algorithm when applied to challenging acoustic data. Our experiments evaluated the effectiveness of different data synthesis strategies and the comprehensive learning-from-synthesis algorithm, which potentially can be applied to artificial intelligence applications such as computer vision, speech recognition, signal processing, and language processing.

### ACKNOWLEDGMENTS

We thank the anonymous reviewers whose poignant comments contributed to improving this article. Thanks to John A. Hildebrand and Simone Baumann-Pickering of Scripps Institution of Oceanography and Melissa S. Soldevilla of the National Oceanic and Atmospheric Administration (NOAA) for providing the acoustic data to the DCLDE 2011 organizing committee. We appreciate the effort of Shannon Rankin and Yvonne Barkley of NOAA in producing portions of the DCLDE 2011 annotations. We thank Michael Weise of the US Office of Naval Research for support (N000141712867).

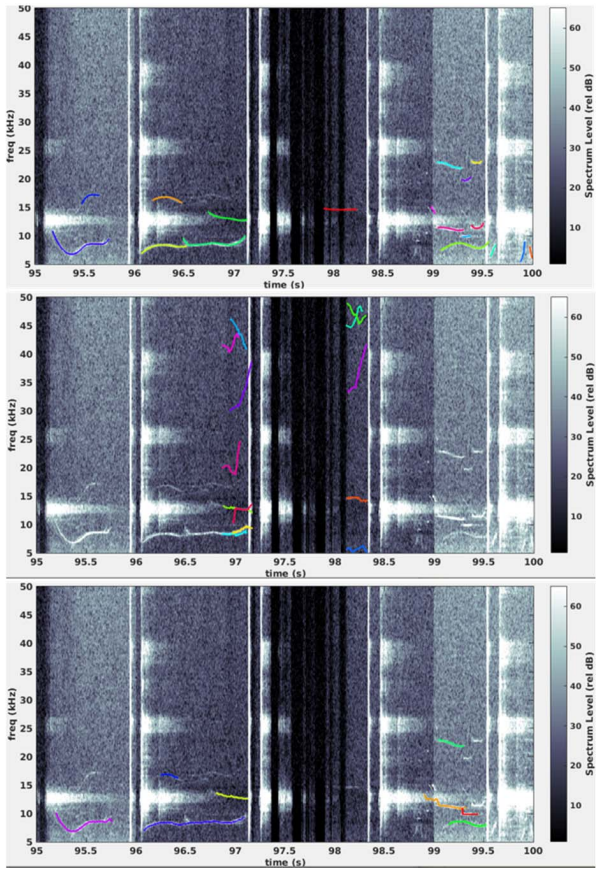

Fig. 7 Whistles, annotations, and predictions in the presence of an echosounder (repeating signal with broad bandwidth). Whistles are overlaid with randomly colored annotations. Annotations were produced by experienced human analysts (top row), predictions from the graph search method [8] (middle row), and predictions from the WGT method, which replaces the graph search signal processing with confidence maps (bottom row).

acoustic clutter. Our deep learning method of identifying whistle energy within a spectrogram had greater precision and recall than standard peak selection mechanisms such as the one used in the graph algorithm. The original graph-based method produced extraneous detections due to false positives in the peak prediction and missed a number of the whistles that the deep whistle contour detector predicted.

**Discussion.** Although whistle contours have complex patterns, the local structure of most whistles is relatively simple. The inner layers of the network may have implicitly captured the representative atoms of a dictionary of whistle